\documentclass[preprint,showpacs,preprintnumbers,amsmath,amssymb,pra,superscriptaddress]{revtex4}

\usepackage[dvips]{graphicx} 
\usepackage{amsfonts}
\usepackage{amscd}
\usepackage{amsmath}    

\begin{document}


\title{Decoy state quantum key distribution with two-way classical post-processing}

\author{Xiongfeng Ma}
 \email{xima@physics.utoronto.ca}
\affiliation{%
Center for Quantum Information and Quantum Control,\\
Department of Physics and Department of Electrical \& Computer Engineering,\\
University of Toronto, Toronto,  Ontario, Canada\\
}%

\author{Chi-Hang Fred Fung}%
 \email{cffung@comm.utoronto.ca}
\affiliation{%
Center for Quantum Information and Quantum Control,\\
Department of Physics and Department of Electrical \& Computer Engineering,\\
University of Toronto, Toronto,  Ontario, Canada\\
}%

\author{Fr\'{e}d\'{e}ric Dupuis}%
 \email{dupuisf@iro.umontreal.ca}
\affiliation{%
D\'{e}partement IRO, Universit\'{e} de Montr\'{e}al, Montr\'{e}al, H3C 3J7 Canada\\
}%

\author{Kai Chen}
\affiliation{%
Center for Quantum Information and Quantum Control,\\
Department of Physics and Department of Electrical \& Computer Engineering,\\
University of Toronto, Toronto,  Ontario, Canada\\
}%

\author{Kiyoshi Tamaki}
 \email{tamaki@will.brl.ntt.co.jp}
\affiliation{%
NTT Basic Research Laboratories, NTT corporation,\\
3-1,Morinosato Wakamiya Atsugi-Shi, Kanagawa, 243-0198, JAPAN\\
}%

\author{Hoi-Kwong Lo}
 \email{hklo@comm.utoronto.ca}
\affiliation{%
Center for Quantum Information and Quantum Control,\\
Department of Physics and Department of Electrical \& Computer Engineering,\\
University of Toronto, Toronto,  Ontario, Canada\\
}%


\begin{abstract}
Decoy states have recently been proposed as a useful method for
substantially improving the performance of quantum key distribution
protocols when a coherent state source is used.
Previously, data post-processing schemes based on one-way classical
communications were considered for use with decoy states.
In this paper, we develop two data post-processing schemes for the
decoy-state method using two-way classical communications. 
Our numerical simulation (using parameters from a specific QKD
experiment as an example)
results show that our 
scheme is able to extend the maximal secure distance from 142km
(using only one-way classical communications with decoy states) to
181km.
The second scheme is able to achieve a 10\% greater key generation
rate in the whole regime of distances.
\end{abstract}

\maketitle

\section{Introduction}
Quantum key distribution (QKD) allows two users, commonly called
Alice (sender) and Bob (receiver), to communicate in absolute
security in the presence of an eavesdropper, Eve. Unlike classical
cryptography, the security of QKD is based on the fundamental
principles of quantum mechanics, rather than unproven computational
assumptions.

The best-known QKD protocol---the BB84 scheme---was published in
1984 \cite{BB84}. In BB84, Alice sends Bob a sequence of single
photons each of which is randomly prepared in one of two conjugate
bases. Bob measures each photon randomly in one of two conjugate
bases. Alice and Bob then publicly compare the bases and keep only
those results (bits) for which they have used the same bases. They
randomly test a subset of those bits and determine the quantum bit
error rate (QBER). If the QBER is larger than some prescribed value,
they abort the protocol. Otherwise, they proceed to the classical
data post-processing (which consists of error correction and privacy
amplification) and generate a secure key. The security of BB84 has
been rigorously proven in a number of papers
\cite{Securityproofs,LoChau,ShorPreskill}, see also \cite{security}.

The security proof in \cite{ShorPreskill} shows that the BB84
protocol can be successively reduced from a entanglement
distillation protocol (EDP). This idea is relevant to this paper
since our data post-processing schemes are based on EDPs. We remark
that EDPs were first discussed in \cite{BBPS}, that its relevance to
the security of QKD was emphasized in \cite{DEJM}, and that this
connection was established rigorously in \cite{LoChau}. The EDPs
proposed earlier use local operations and one-way classical
communications (1-LOCC). Later, Gottesman and Lo provided security
proofs of standard quantum key distribution schemes by using a EDP
with local operations and two-way classical communications (2-LOCC)
\cite{GL}. They showed that BB84 using 2-LOCC can tolerate a higher
bit error rate than 1-LOCC (see also \cite{Chau2002}). On the other hand, Gerd, Vollbrecht and
Verstraete also proposed another EDP that uses a 2-LOCC based
recurrence scheme \cite{GVV}.
Although their scheme was originally proposed as an EDP, we will use it
here in a QKD to increase the key generation rate.
[It should be noted that the EDP approach is only one of the several approaches to security proofs of QKD.
Other useful approaches to security proof can be based on, for example, communication complexity \cite{Ben-Or2002},
quantum memory \cite{Renner2004, Christandl2004}, or direct information-theoretic argument \cite{Renner2005}.]
Recently, it has been demonstrated \cite{HLLO} rigorously that one can generate a long secure key even when
the amount of distillable entanglement in a quantum state is arbitrarily small.
In other words, secure key generation is strictly weaker than entanglement distillation.
Recently, the universal composability of quantum key distribution has been proven in \cite{Ben-Or2004}.

In summary, QKD is secure in theory. Much of the interest in QKD is
due to its potential in near-term real-life applications. Indeed,
commercial optical-fiber-based quantum cryptographic products are
already on the market \cite{MagiQidQuantique}.

Meanwhile, experimentalists have done many QKD experiments, such as
\cite{Bennett1992} and \cite{GYS,NEC}.
The key issue in QKD experiments is whether they are really secure.
Standard security proofs are often based on perfect devices,
such as perfect single photon sources.
All devices are imperfect in real
implementations, such as imperfect single photon sources and highly
lossy channels.
It is thus important to study the security of QKD
with imperfect devices. Substantial progress has been made in the
subject \cite{ILM,GLLP,Koashi2004}.

Unfortunately, with the method in GLLP \cite{GLLP}, QKD can only be
proven to be secure at very limited key generation rates and
distances. It came as a big surprise that a simple solution to the
problem --- the decoy state method --- actually exists. The decoy
method was first discovered by Hwang \cite{HwangDecoy}, and made
rigorous by our group \cite{Decoy, Practical}, and also
\cite{WangDecoy, Harrington}. In addition, our group demonstrated
the first experimental implementation of a QKD protocol using one
decoy state~\cite{Yi2005}.

The usefulness of decoy state protocols over non-decoy-state protocols have previously been demonstrated \cite{Decoy, Practical},
within the context of 1-LOCCs, for an imperfect source.
Since 2-LOCCs are known to be superior to 1-LOCCs for a perfect source \cite{GL,Chau2002,GVV}, it would be interesting to study the usefulness of decoy state protocols with 2-LOCCs, for an imperfect source.
This is the main goal of this paper.
Indeed, as we will show,
decoy state protocols with 2-LOCCs are superior to decoy state protocols with only 1-LOCCs in realistic situations.
Specifically,
in this paper, we develop two data post-processing schemes for the
decoy method of \cite{Decoy,Practical} by applying two 2-LOCC EDPs,
Gottesman-Lo EDP \cite{GL} and the recurrence scheme \cite{GVV}.
Both methods are superior to the random hashing 1-LOCC EDP (for the rest of the paper, we will simply call it as the 1-LOCC EDP)
in two different aspects in
the case of ideal devices; the Gottesman-Lo EDP was shown to be able
to achieve a higher tolerable bit error rate, while the recurrence
method was shown to be able to achieve a higher key generation rate.
We will show in this paper that the same conclusion holds in the
case of imperfect devices.
In particular, depending on the distance in a QKD experiment, one can use our Gottesman-Lo EDP based data post-processing scheme in the long distance region or
our recurrence based data post-processing scheme in the short distance region to increase the key generation rate.

We note that a recent and independent analysis of combining B steps
with GLLP and decoy states is given in \cite{Khalique2006}. Their
data post-processing scheme is the same as our first scheme, which
is aimed at increasing the maximal secure distance. On the other
hand, in this paper, we also propose the second scheme, which is
aimed at increasing the key rate at short distances.

The organization of this paper is as follows:
We first review entanglement distillation in Section~\ref{review-EDP} and
some existing techniques for realistic QKD in Section~\ref{review-realistic-QKD}.
We then investigate the tolerable error rates, the upper bounds of
secure distance and key generation rate in Section~\ref{Bounds}.
Sections \ref{BDecoy} and \ref{Recurrence} contain the main results
of the paper.
Specifically, we
develop two data post-processing schemes, one with the B steps from
Gottesman and Lo \cite{GL} (see Section \ref{BDecoy}), and the other
with recurrence (see Section \ref{Recurrence}). Our simulation
(based on the GYS experiment \cite{GYS}) shows that with B steps
from Gottesman-Lo EDP, the maximal secure distance can be extended
to 180km compared with 140km with 1-LOCC, and the key generation
rate increased by more than 10\% in the whole regime of distances.
With our QKD model, we also consider statistical fluctuations on the
estimated parameters when the data has finite length (see Section
\ref{StaFlu}). The result shows that the B step can extend the
maximal secure distance and the recurrence can raise the key generation
rate. Although, in this paper, we focus on the BB84 protocol, our
schemes can be applied to other QKD protocols as well.

%

\section{Review of entanglement distillation\label{review-EDP}}
In this section, we review Shor-Preskill's security proof
of QKD and two EDPs based on 2-LOCC (Gottesman-Lo EDP and recurrence
EDP) assuming that ideal single-photon sources are used.
In Sections~\ref{BDecoy} and
\ref{Recurrence}, we generalize these two schemes for realistic
setups.

The idea of the Shor-Preskill \cite{ShorPreskill} security proof of
QKD is to apply an EDP to show that the leaked information about the
final key is negligible. Here we will explain how to analyze the
security of EDP-based QKD.

In the EDP-based QKD protocol, Alice creates $n+m$ pairs of qubits,
each in the state
$$|\psi\rangle=\frac{1}{\sqrt2}(|00\rangle+|11\rangle),$$
the eigenstate with eigenvalue $1$ of the two commuting operators
$X\bigotimes X$ and $Z\bigotimes Z$, where
$$
X=\begin{pmatrix}
    0 & 1\\
    1 & 0\\
    \end{pmatrix},
Z=\begin{pmatrix}
    1 & 0\\
    0 & -1\\
    \end{pmatrix}
    $$
are the Pauli operators. Then she sends half of each pair to Bob.
Alice and Bob sacrifice $m$ randomly selected pairs to test the
error rates in the $X$ and $Z$ bases by measuring $X\bigotimes X$
and $Z\bigotimes Z$. If the error rates are too high, they abort the
protocol. Otherwise, they conduct the EDP, extracting $k$
high-fidelity pairs from the $n$ noisy pairs. Finally, Alice and Bob
both measure $Z$ on each of these pairs, producing a \textit{k-bit}
shared random key about which Eve has negligible information. The
protocol is secure because the EDP removes Eve's entanglement with
the pairs, leaving her negligible knowledge about the outcome of the
measurements by Alice and Bob.

In the EDP, after the qubits' transmission, Alice and Bob will share
the state with density matrix,
\begin{equation} \label{EDP:densitymatrix}
\begin{aligned}
\rho=\begin{pmatrix}
    q_{00} & \times & \times & \times\\
    \times & q_{10} & \times & \times \\
    \times & \times & q_{11} & \times \\
    \times & \times & \times & q_{01} \\
    \end{pmatrix},
\end{aligned}
\end{equation}
normalized with $q_{00}+q_{10}+q_{11}+q_{01}=1$.
Here $\times$'s denote arbitrary numbers, all of which are not
necessary the same,
and the density matrix is in
the Bell basis:
$$
\begin{aligned}
|\psi_{00}\rangle&=\frac{1}{\sqrt2}(|00\rangle+|11\rangle)\\
|\psi_{10}\rangle&=\frac{1}{\sqrt2}(|01\rangle+|10\rangle)\\
|\psi_{11}\rangle&=\frac{1}{\sqrt2}(|01\rangle-|10\rangle)\\
|\psi_{01}\rangle&=\frac{1}{\sqrt2}(|00\rangle-|11\rangle).\\
\end{aligned}
$$
Since all EDPs we consider in this paper do not make use of the
off-diagonal elements in Eq.~\eqref{EDP:densitymatrix} to extract
entanglement,
it is sufficient to characterize the
density matrix
by only the diagonal elements
$(q_{00}, q_{10}, q_{11},
q_{01})$.
In fact, any shared state can always be transformed into a diagonal
form by local operations and classical communications \cite{BDSW}.
The density matrix is now a classical mixture of the Bell states
$\psi_{ij}$ with probabilities $q_{ij}$. Therefore, the bit and
phase error rates are given by
\begin{equation}
\begin{aligned}
\delta_b=q_{10}+q_{11} \\
\delta_p=q_{11}+q_{01}.  \\
\end{aligned}
\end{equation}

A QKD protocol based on a Calderbank-Shor-Steane (CSS) \cite{CSS} EDP can be reduced to a
``prepare-and-measure" protocol (BB84) \cite{ShorPreskill}.
That is to say, CSS codes correct bit errors and phase errors separately,
which respectively turn out to be the bit error correction and
privacy amplification in the context of QKD \cite{ShorPreskill}.
Thus the key rate of this 1-LOCC based data post-processing scheme is given by
\cite{ShorPreskill,Koashi2005},
\begin{equation}\label{EDP:CSSrate}
R_{CSS}=q\left[1-H_2(\delta_b)-H_2(\delta_p)\right],
\end{equation}
where
$q$ depends on the implementation (1/2 for the BB84
protocol, because half the time Alice and Bob bases are not
compatible, and if we use the efficient BB84 protocol
\cite{EffBB84}, we can have $q\approx1$.),
$\delta_b$ and $\delta_p$ are the bit flip error rate and the
phase flip error rate, and $H_2(x)$ is the binary entropy function,
$$
H_2(x)=-x\log_2(x)-(1-x)\log_2(1-x).
$$

In summary, there are two main parts of EDP, bit flip error
correction (for error correction) and phase flip error correction
(for privacy amplification). These two steps can be understood as
follows. First Alice and Bob apply error correction,
after which they share the same key strings but Eve may still keep
some information about the key. Alice and Bob then perform the
privacy amplification to expunge Eve's information from the key.
We remark that the key generation rate achieved by Eq.~\eqref{EDP:CSSrate} requires only 1-LOCC.


\subsection{Gottesman-Lo EDP} \label{BP}
Gottesman and Lo \cite{GL} introduced an EDP based on 2-LOCC for use with QKD
and showed the it can tolerate a higher bit error rate than 1-LOCC based EDP's.
B and P steps are two primitives in the Gottesman-Lo EDP, and the
EDP consists of executing a sequence of B and/or P steps, followed
by random hashing. The random hashing part is a one-way
EDP. The main objective for extra B and P steps is reduce the bit
and/or phase error rates of qubits so that the random hashing can
work to extract secure keys.
This is the reason why the Gottesman-Lo EDP is able to tolerate a
higher initial bit error rate than one-way EDPs. The definitions of
B and P steps are as follows.

\begin{figure}[hbt]
\centering \resizebox{8cm}{!}{\includegraphics{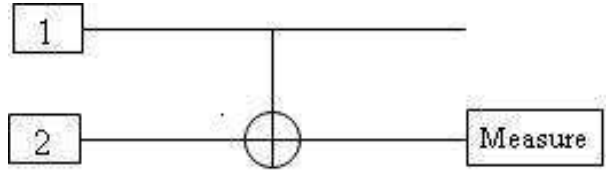}}
\caption{Alice and Bob choose two half EPR pairs and input the
quantum circuit as shown above. They discard both control and target
qubits if they disagree on the outcome of measurement on the target
qubits. On the other hand, they keep the control qubits as surviving
qubits if their measurement outcomes agree.
}%
\label{Fig:BStepCircuit}
\end{figure}

\textbf{Definition of B step \cite{GL}:} (Figure
\ref{Fig:BStepCircuit}) After randomly permuting all the EPR pairs,
Alice and Bob perform a bilateral XOR (BXOR) between pairs of EPR pairs and
measure the target qubits in $Z$ basis. This effectively measures
the operator $Z\bigotimes Z$ for each of Alice and Bob, and detects
the presence of a single bit flip error. If Alice and Bob's
measurement outcomes disagree, they discard the remaining EPR pair,
otherwise, they keep the control qubit.

Since the B step only involves the measurement of $Z\bigotimes Z$,
it can be used in the prepare-and-measure protocol, BB84.
Classically, the B step simply involves random pairing of the key
bits, say $x_1, x_2$ on Alice's side and $y_1, y_2$ on Bob's side
and the computation of the parity of each pair of bits, $x_1 \oplus
x_2$ and $y_1 \oplus y_2$. Both Alice and Bob announce the parities.
If their parities are the same, they keep $x_1$ and $y_1$;
otherwise, they discard $x_1$, $x_2$, $y_1$ and $y_2$. We can see
that the B step is very simple to implement in data post-processing.

Suppose Alice and Bob input: a control qubit $(q_{00}^C, q_{10}^C,
q_{11}^C, q_{01}^C)$ and a target qubit $(q_{00}^T, q_{10}^T,
q_{11}^T, q_{01}^T)$ with bit error rates $\delta_b^C$ and
$\delta_p^C$ and phase error rates $\delta_b^T$ and $\delta_p^T$,
respectively. After one B step, the survival probability $p_{S}$ is
given by,
\begin{equation}\label{oneBpS}
\begin{aligned}
p_{S} &= (q_{00}^C+q_{01}^C)(q_{00}^T+q_{01}^T) + (q_{10}^C+q_{11}^C)(q_{10}^T+q_{11}^T)\\
      &= (1-\delta_b^C)(1-\delta_b^T) + \delta_b^C\delta_b^T,
\end{aligned}
\end{equation}
and the density matrix $(q_{00}', q_{10}', q_{11}', q_{01}')$ of
output control qubit is given by
\begin{equation}\label{Twoway:oneBmatrix}
\begin{aligned}
q_{00}' &= \frac{q_{00}^Cq_{00}^T + q_{01}^Cq_{01}^T}{p_{S}}\\
q_{10}' &= \frac{q_{10}^Cq_{10}^T + q_{11}^Cq_{11}^T}{p_{S}}\\
q_{11}' &= \frac{q_{10}^Cq_{11}^T + q_{11}^Cq_{10}^T}{p_{S}}\\
q_{01}' &= \frac{q_{00}^Cq_{01}^T + q_{01}^Cq_{00}^T}{p_{S}}.
\end{aligned}
\end{equation}
Eqs.~\eqref{Twoway:oneBmatrix} can be derived from TABLE II of
\cite{BDSW}. The corresponding bit error rate $\delta_b$ and phase
error rate $\delta_p$ can be obtained from
Eq.~\eqref{Twoway:oneBmatrix} by
\begin{equation}\label{Twoway:AfterBstepErr}
\begin{aligned}
\delta_b' &= q_{10}'+q_{11}' = \frac{\delta_b^C\delta_b^T}{p_S}\\
\delta_p' &= q_{11}'+q_{01}'.\\
\end{aligned}
\end{equation}

\textbf{Definition of P step \cite{GL}:} Randomly permute all the
EPR pairs. Afterwards, group the EPR pairs into sets of three, and
measure $X_1X_2$ and $X_1X_3$ on each set (for both Alice and Bob).
This can be done (for instance) by performing a Hadamard transform,
two bilateral XORs, measurement of the last two EPR pairs, and a
final Hadamard transform. If Alice and Bob disagree on one
measurement, Bob concludes the phase error was probably on one of
the EPR pairs which was measured and does nothing; if both
measurements disagree for Alice and Bob, Bob assumes the phase error
was on the surviving EPR pair and corrects it by performing a $Z$
operation.


Without a quantum computer, Alice and Bob are not able to perform
the P steps,
so the EDP cannot depend on the results of P steps. When the P step
is implemented classically in BB84, the phase errors are not
detected or corrected (i.e. the phase flip operation $Z$ is not
applied).
The P step then will be reduced to: Alice and Bob randomly form
trios of the remaining qubits and compute the parity of each trio,
say $x_1 \oplus x_2 \oplus x_3$ by Alice and $y_1 \oplus y_2 \oplus
y_3$ by Bob. They now regard those parities as their new bits for
further processing.

Since before P steps, Alice and Bob will do random permutation, for
simplicity, we assume the input three qubits have the same density
matrix: $(q_{00}, q_{10}, q_{11}, q_{01})$. After one P step, the
density matrix $(q_{00}', q_{10}', q_{11}', q_{01}')$ of the output
qubit is given by
\begin{equation}\label{Twoway:onePmatrix}
\begin{aligned}
q_{00}' &= q_{00}^3+3q_{00}^2q_{01}+3q_{10}^2(q_{00}+q_{01})+6q_{00}q_{10}q_{11}\\
q_{10}' &= q_{10}^3+3q_{10}^2q_{11}+3q_{00}^2(q_{10}+q_{11})+6q_{00}q_{10}q_{01}\\
q_{11}' &= q_{11}^3+3q_{10}q_{11}^2+3q_{01}^2(q_{10}+q_{11})+6q_{00}q_{11}q_{01}\\
q_{01}' &= q_{01}^3+3q_{00}q_{01}^2+3q_{11}^2(q_{00}+q_{01})+6q_{10}q_{11}q_{01},\\
\end{aligned}
\end{equation}
which is given in Appendix C of \cite{GL}. So the bit error rate and
phase error rate will given by
\begin{equation}\label{Twoway:AfterPstepErr}
\begin{aligned}
\delta_b' &= q_{10}'+q_{11}' = 3\delta_b(1-\delta_b)^2+\delta_b^3 \\
\delta_p' &= q_{11}'+q_{01}' = 3\delta_p^2(1-\delta_p)+\delta_p^3. \\
\end{aligned}
\end{equation}

Here we emphasize that the B and P steps are important elements of
the Gottesman-Lo EDP. After B and P steps, the Gottesman-Lo EDP will
be the same as the regular 1-LOCC EDP.

\subsection{Recurrence EDP scheme} \label{Rec:Subsection:Rec}
Here we review another two-way EDP, the recurrence scheme
\cite{GVV}.
Similar to the B step in Gotttesman-Lo EDP, the recurrence scheme reduces the bit error rate of the EPR pairs before passing them to the 1-LOCC based random hashing for the distillation of maximally-entangled EPR pairs.
However, there are two main differences between these two EDP schemes.
The first is how the bit error syndrome of a target EPR pair in a bilateral XOR operation is learned.
In Gotttesman-Lo EDP, Alice and Bob simply measure the target EPR pair in the $Z$ basis and compare their results to learn the bit error syndrome
(see Figure \ref{Fig:BStepCircuit}).
In the recurrence scheme,
Alice and Bob group the bit error syndromes of all target EPR pairs together and learn all the syndromes using random hashing.
The second difference is that the recurrence scheme requires some extra maximally-entangled EPR pairs to begin with for learning the bit error syndromes,
whereas no such extra pairs are required in the Gotttesman-Lo EDP.
We note that the recurrence methods have been
studied in various papers, such as
\cite{DEJM,recurrence2,recurrence3,recurrence4}.

The steps for the recurrence protocol are as follows:
\begin{enumerate}
\item
Alice and Bob perform BXOR using two noisy EPR pairs as the sources
and one perfect maximally-entangled EPR pair as the target.

\item
They do random hashing on the target EPR pairs to learn the parities of noisy EPR pairs.
Note that only a portion of the target EPR pairs have to be measured in order to learn all the parities.
This is different from the B step approach.

\item
They do error correction and privacy amplification separately for even-parity EPR pairs and odd-parity EPR pairs.

\end{enumerate}

The key generation rate using the recurrence EDP with a single-photon source is
given by
\begin{equation} \label{Rec:Residue0a}
\begin{aligned}
R = q \left[ -\frac12H_2(p_S)
-\frac12p_SH_2(\frac{\delta_b^C\delta_b^T}{p_S}) + K \right]
\end{aligned}
\end{equation}
where
$q$ is defined similarly as in Eq.\eqref{EDP:CSSrate},
$p_S$ (given in Eq.~\eqref{Rec:Survival}) is the probability of getting even parity, and
$\delta_b^C$($\delta_b^T$) is the bit error rate of the control (target) EPR pair.
Here,
the first term in the bracket corresponds to the extra perfect EPR pairs borrowed,
the second term corresponds to error correction, and
the third term $K$ (given in Eq.\eqref{Rec:PriRes2}) corresponds to privacy amplification.
In Appendix~\ref{App:review-recurrence}, we review the recurrence EDP in detail and develop the
key rate formula.

\section{Review of realistic QKD\label{review-realistic-QKD}}
In this section, we set up a model for realistic QKD, and
review the idea of GLLP and decoy-state QKD.

\subsection{Realistic QKD setup} \label{Model}
In this section, we present a commonly used fiber-based QKD
system model. All later simulations of QKD are based on this model.
In order to describe a real-world QKD system, we need to model the
source, transmission and detection. Here we consider a widely used
QKD setup model with polarization coding \cite{Lutkenhaus}, see also
\cite{Practical}.

\textbf{Source:} The laser source used in the QKD experiment can be
modeled as a weak coherent state. Assuming that the phase of each
pulse is totally randomized, the photon number of each pulse follows
a Poisson distribution with a parameter $\mu$ as its expected photon
number set by Alice. Thus, the density matrix of the state emitted
by Alice is given by
\begin{equation}\label{Model:AliceState}
\rho_A=\sum^{\infty}_{i=0}\frac{\mu^i}{i!}\,e^{-\mu}\,
|i\rangle\langle i|,
\end{equation}
where $|0\rangle\langle 0|$ is \textit{vacuum state} and
$|i\rangle\langle i|$ is the density matrix of the $i$-photon state
for $i=1,2\cdots$. The states with only one photon ($i=1$) are
normally called \emph{single photon states}. The states with more
than one photon ($i\ge2$), on the other hand, are called
\textit{multi photon states}. Here, we assume Eve receives all the
pulses sent by Alice. Eve performs some arbitrary operations and
sends either a vacuum or a qubit to Bob.
This is the squash operation introduced in GLLP \cite{GLLP}.
Consequently, we denote the qubits coming from these three states as
{vacuum qubits}, {single photon qubits} and {multi photon qubits}.

A vacuum qubit is
a qubit sent by Eve when Alice sent a vacuum state.
(In the case without Eve's presence, it is
some random qubit 
stemmed from the dark counts of Bob's detector or other background contributions.)
Thus, it does not contribute to the key generation.
Due to photon-number splitting
(PNS) attacks \cite{PNS,PNS1st,BLMS,LJPNS}, multi photon states are
not secure for the BB84 protocol. Here is a key observation of this
QKD model: \emph{the final secure key can only be extracted from
single photon qubits}\footnote{That only single photon qubits contribute to the secure key is only true for a security proof based on the EDP approach (which is what we use in this paper).  It may not be true in other approaches, e.g., the communication complexity approach, as noted in Ref. \cite{Tamaki}.}. Besides BB84, this is true for most present
QKD protocols, such as E91 \cite{E91}, B92 \cite{B92} and the
six-state \cite{sixstate} scheme. One exception is the SARG04
protocol \cite{SARG04}, in which two-photon states can also
contribute to the secure key generation rate \cite{Tamaki,Fung}.

\textbf{Transmission:} For optical fiber based QKD systems, the
losses in the quantum channel can be derived from the loss
coefficient $\alpha$ measured in dB/km and the length of the fiber
$l$ in km. The overall transmittance is given by
\begin{equation}\label{Model:Eta}
\eta=\eta_{Bob}10^{-\frac{\alpha l}{10}}.
\end{equation}
where $\eta_{Bob}$ denotes for the transmittance in Bob's side,
including the internal transmittance of optical components and
detector efficiency. Here we assume a threshold single photon
detector on Bob's side. That is to say, we assume that Bob's
detector can tell whether there is a click or not.
However, it cannot tell the actual photon number of the received
signal, if it contains at least one photon.

It is reasonable to assume independence between the behaviors of the
$i$ photons in $i$-photon states. Therefore the transmittance of
\emph{$i$-photon} state $\eta_i$ with respect to a threshold
detector is given by
\begin{equation}\label{Model:etai}
\eta_i=1-(1-\eta)^i
\end{equation}
for $i=0,1,2,\cdots$.

\textbf{Yield:} Define $Y_i$ to be the yield of an $i$-photon state,
i.e., the conditional probability of a detection event at Bob's side
given that Alice sends out an $i$-photon state. Note that $Y_0$ is
the background rate which includes detector dark counts and other
background contributions such as the stray light from timing pulses.

The yield of the $i$-photon states $Y_i$ mainly comes from two
parts, the background and the true signal. Assuming that the
background counts are independent of the signal photon detection,
then $Y_i$ is given by
\begin{equation}\label{Model:Yi}
\begin{aligned}
Y_i &= Y_0 + \eta_i - Y_0\eta_i \\
    &\cong Y_0 + \eta_i.
\end{aligned}
\end{equation}
Here we assume $Y_0$ (typically $10^{-5}$) and $\eta$ (typically
$10^{-3}$) are small.

The {\it gain} of $i$-photon states $Q_i$ is given by
\begin{equation}\label{Model:Qi}
\begin{aligned}
Q_i &= Y_i\frac{\mu^i}{i!}e^{-\mu}.
\end{aligned}
\end{equation}
The gain $Q_i$ is the product of the probability of Alice sending
out an $i$-photon state (follows Poisson distribution) and the
conditional probability of Alice's $i$-photon state (and background)
that will lead to a detection event in Bob's detector.

\textbf{Quantum Bit Error Rate (QBER):} The error rate of $i$-photon
states $e_i$ is given by
\begin{equation}\label{Model:ei}
e_i = \frac{e_0 Y_0 + e_{d}\eta_i}{Y_i}
\end{equation}
where $e_{d}$ is the probability that a photon hit the erroneous
detector. $e_{d}$ characterizes the alignment and stability of the
optical system. Experimentally, even at distances as long as 120km,
$e_{d}$ is independent of the distance \cite{GYS}. In what follows,
we will also assume that $e_{d}$ is independent of the transmission
distance
We will assume that the background is random. Thus the error rate of
the background is $e_0=\frac12$. Note that Eqs.~\eqref{Model:etai},
\eqref{Model:Yi}, \eqref{Model:Qi} and \eqref{Model:ei} are
satisfied for all $i=0,1,2,\cdots$.

The overall gain is given by
\begin{equation}\label{Model:Gain}
\begin{aligned}
Q_{\mu} &= \sum_{i=0}^{\infty} Y_i\frac{\mu^i}{i!}e^{-\mu}. \\
\end{aligned}
\end{equation}

The overall QBER is given by
\begin{equation}\label{Model:QBER}
\begin{aligned}
E_{\mu} &= \frac{1}{Q_{\mu}} \sum_{i=0}^{\infty} e_iY_i\frac{\mu^i}{i!}e^{-\mu}. \\
\end{aligned}
\end{equation}

Without Eve, a normal QKD transmission will give
\begin{equation}\label{Model:WithoutEve}
\begin{aligned}
Q_{\mu} &= Y_0 + 1-e^{-\eta\mu} \\
E_{\mu}Q_{\mu} &= e_0 Y_0 + e_{d}(1-e^{-\eta\mu}).
\end{aligned}
\end{equation}

\subsection{GLLP idea} \label{GLLP}
We review the idea of GLLP's \cite{GLLP} briefly here. GLLP
gives a security proof of BB84 QKD when imperfect devices (such as
imperfect single photon sources) are used. There are two kind of
qubits discussed in GLLP, tagged qubits and untagged qubits.
Tagged qubits are those that have their basis information revealed to Eve,
i.e. tagged qubits are not secure for QKD.
On the other hand, untagged qubits are secure for QKD.
In BB84, qubits coming from single-photon states are untagged while
those from multi-photon states are tagged because Eve, for instance, can
perform PNS attacks \cite{PNS,PNS1st,BLMS,LJPNS} to the multi-photon states
to acquire their basis information.
The essential idea of
GLLP is that Alice and Bob can apply privacy amplification to tagged
and untagged qubits separately.
Note that the idea of tagged state was (perhaps implicitly)
introduced by \cite{ILM}.

The data post-processing of GLLP is performed as follows. First,
Alice and Bob apply error correction to all qubits, sacrificing a
fraction $H_2(\delta)$ of the key, which is represented in the first
term of Eq.~\eqref{GLLP:GLLPex}. Secondly, in principle, Alice and
Bob can distinguish the tagged and untagged qubits, so they can
apply the privacy amplification on the tagged state and untagged
state separately. One can imagine executing privacy amplification on
two different strings, the qubits $s_{tagged}$ and $s_{untagged}$
arising from the tagged qubits and the untagged qubits respectively.
Since the privacy amplification is linear (the private key can be
computed by applying the $C_2$ parity check matrix to the qubit
string), the key obtained is the bitwise $XOR$
$$
s_{untagged}\oplus s_{tagged}
$$
of keys that could be obtained from the tagged and untagged qubits
separately \cite{GLLP}. If $s_{untagged}$ is private and random,
then it doesn't matter if Eve knows anything about $s_{tagged}$ ---
the sum will be still private and random. Thus, one only needs to
apply privacy amplification to the untagged bits alone.

We define the residue of data post-processing to be the ratio of the
final key length to the sifted key length (in an asymptotic sense).
The residue of this data post-processing scheme is given by
\begin{equation}\label{GLLP:GLLPex}
r_{GLLP}=\max\{-f(\delta)H_2(\delta)+\Omega[1-H_2(\delta_p)],0\}
\end{equation}
where $\delta$ is the overall quantum bit error rate (QBER),
$\Omega$ is the fraction of untagged qubits ($\Omega=1-\Delta$,
where $\Delta$ is the fraction of tagged qubits defined in GLLP
\cite{GLLP}), $\delta_p$ is the phase error rate of the untagged
qubits, $f(\cdot)$ is the error correction efficiency as a function
of error rate \cite{BS}, normally $f(x)\ge1$ with Shannon limit
$f(x)=1$, and $H_2(x)$ is binary entropy function.

We can further extend GLLP's idea to the case of more than two
classes of qubits, i.e. several kinds of qubits with flag $g$, which
generalizes the concept of tagged and untagged qubits.
The procedure of data post-processing is similar, do the overall
error correction first and then apply the privacy amplification to
each case. So the privacy amplification part can be written as
\begin{equation}\label{GLLP:Priex}
\sum_g \Omega^gH_2(\delta_p^g)
\end{equation}
where one needs to sum over all cases with flag $g$, $\Omega_g$ is
the probability of the case with flag $g$ and $\sum_g \Omega^g=1$,
and $\delta_p^g$ is the phase error rate of the state with flag $g$.
At last, the residue of data post-processing is given by
\begin{equation}\label{GLLP:RateGeneral}
r=\max\{-f(\delta)H_2(\delta)+\sum_g
\Omega^g[1-H_2(\delta_p^g)],0\}.
\end{equation}

Apply the QKD model described in Section \ref{Model} here,
$\delta=E_{\mu}$ and the key generation rate is given by
\begin{equation}\label{GLLP:KeyRate}
R = q\cdot Q_{\mu}\cdot r,
\end{equation}
where $Q_{\mu}$ and $E_{\mu}$ is the gain and QBER of the signal
state, and
$q$ is defined similarly as in Eq.\eqref{EDP:CSSrate}.

\subsection{Decoy states} \label{Decoy}
For BB84, the single photon state is the only source of final secure
keys, i.e.~the untagged qubits come from single photon qubits. So
the fraction and error rate of untagged qubit are given by
\begin{equation}\label{Decoy:Untag}
\begin{aligned}
\Omega&=Q_1/Q_\mu \\
\delta_p&=e_1,
\end{aligned}
\end{equation}
where $Q_1$ is given in Eq.~\eqref{Model:Qi}, $e_1$ is given by
Eq.~\eqref{Model:ei}, and $Q_\mu$ is given by
Eq.~\eqref{Model:Gain}. By substituting Eq.~\eqref{GLLP:GLLPex}, we
can rewrite Eq.~\eqref{GLLP:KeyRate} into
\begin{equation}\label{Decoy:KeyRate}
\begin{aligned}
R &= q\cdot r\cdot Q_{\mu} \\
&\ge q\cdot\{-Q_{\mu}f(E_{\mu})H_2(E_{\mu}) +Q_1[1-H_2(e_1)]\},
\\
\end{aligned}
\end{equation}
which is given in Eq.~(11) of \cite{Decoy}.

$Q_\mu$ and $E_\mu$ can be measured directly from experiment. The
question is how to estimate $Q_1$ and $e_1$ accurately? In
principle, Eve can perform non-demolition photon number measurement
on the qubits and she may change the yields ($Y_i$ in Subsection
\ref{Model}) of the qubits depending on the measurement outcomes.
That is, the yields of qubits, in general, may depend on the photon
number. Moreover, Eve can adjust the error rates as she wishes.

The key idea of decoy states is that, instead of just using one
coherent state for key transmission, Alice and Bob choose some decoy
states with different expected photon numbers ($\mu$ in Subsection
\ref{Model}) to test the channel transmittance and error rate. We
emphasize here that decoy states have exactly the same properties
other than average photon numbers, so that there is no way for Eve
to discriminate between the signal states and decoy states before
Alice publicly announce them. Consequently, we have
$Y_i(decoy)=Y_i(signal)$ and $e_i(decoy)=e_i(signal)$.

Specifically, the overall gain and the overall QBER in
Eq.~\eqref{Model:Gain} and Eq.~\eqref{Model:QBER} can be estimated
for a fixed $\mu$ in the experiment by Alice and Bob. By changing
$\mu$ over many values, a set of linear equations in the form of
Eq.~\eqref{Model:Gain} and Eq.~\eqref{Model:QBER} with unknowns
$Y_i$'s and $e_i$'s are obtained. Thus, Alice and Bob can easily
solve for $Y_i$'s and $e_i$'s from these equations. For BB84, they
are only interested in $Y_1$ and $e_1$. With the decoy state, Alice
and Bob can estimate the yields and error rates of single photon
states ($Y_1$ and $e_1$) accurately.

Here we will briefly review the results of decoy state protocols.
Details can be seen in \cite{Decoy} and \cite{Practical}. In the
\emph{asymptotic} decoy states case \cite{Decoy}, we assume that
infinite decoy states are employed by Alice and Bob, so they can
solve the infinite number of linear equations in the form of
Eq.~\eqref{Model:Gain} and Eq.~\eqref{Model:QBER} to get all values
of $\{Y_i\}$ and $\{e_i\}$ accurately. In the simulation, we will
simply take the value of Eq.~\eqref{Model:Yi} and
Eq.~\eqref{Model:ei} directly.

In the \emph{practical} case, Alice and Bob only need to use two
decoy states, a vacuum and weak decoy state. Then they can bound
$Y_1$ and $e_1$ by (Eq.~(34) and Eq.~(37) in \cite{Practical})
\begin{equation}\label{Decoy:Y1e1prac}
\begin{aligned}
Y_1 &\ge \frac{\mu}{\mu\nu-\nu^2}(Q_\nu e^{\nu}-Q_\mu
e^\mu\frac{\nu^2}{\mu^2}-\frac{\mu^2-\nu^2}{\mu^2}Y_0) \\
e_1 &\le \frac{E_\nu Q_\nu e^{\nu}-e_0Y_0}{Y_1\nu},
\end{aligned}
\end{equation}
where $\nu$ is the expected photon number of weak decoy state. We
remark that when $\nu\rightarrow0$, Eqs.\eqref{Decoy:Y1e1prac} will
asymptotically approach Eq.~\eqref{Model:Yi} and
Eq.~\eqref{Model:ei}.

\section{Bounds} \label{Bounds}
In QKD experiments, we are interested in maximizing three quantities
-- the tolerable error rates, the key generation rate and the
maximal secure distance. In this section, we will give out these
three bounds due to QKD setup model discussed in Subsection
\ref{Model}.

\subsection{Bounds of error rates} \label{Boundary}
Here, we will consider the bounds of error rates (bit error rate
$\delta_b$ and phase error rate $\delta_p$), assuming a laser source that emits a fixed number of photons in each pulse (e.g. a basis-dependent single-photon source).
The upper bounds can be
derived by considering some simple attacks (such as intercept-resend
attack) and determining the QBER caused by these attacks.
The lower bounds can be determined by the unconditionally security
proof assuming that Eve is performing arbitrary attack allowed by
the law of quantum mechanics and Alice and Bob employ some special
data post-processing schemes (such as Gottesman-Lo EDP described in
Subsection \ref{BP}). One lower bound, obtained by considering
Gottesman-Lo EDP, is $18.9\%$ \cite{GL}. For BB84, an upper bound,
obtained by considering an intercept-resend attack, is $25\%$.

Here, we consider the lower bound in a general setting that the
error rates are characterized by $(\delta_b,\delta_p)$. In general,
the bit error rate $\delta_b$ can be measured by error testing, but
the phase error rate $\delta_p$ cannot be directly observed from the
QKD experiment. In order to guarantee the security, Alice and Bob
have to bound $\delta_p$ with the knowledge of $\delta_b$. For BB84 with an ideal single-photon source,
due to the symmetry between the $X$ and $Z$ bases, one can show that the
bit error rate and the phase error rate are the same, i.e.
\begin{equation} \label{Twoway:b=p}
\delta_b=\delta_p.
\end{equation}
In general, for other protocols or with non-ideal sources (including coherent sources), the bit and phase error rates are
different.
For example, even for BB84, when a basis-dependent source is used,
Eq.~\eqref{Twoway:b=p} may not hold.
In this case, according to the
idea of \cite{Koashi2005}, we can show that $\delta_b$ and
$\delta_p$ have the relation of
\begin{equation} \label{Bounds:bp}
F\le\sqrt{(1-\delta_b)(1-\delta_p)}+\sqrt{\delta_b
\delta_p},
\end{equation}
where $F$ is the fidelity between the two states sent by Alice
corresponding to the two bases and is the single parameter that
characterizes the basis dependency of the source. Thus, Alice and
Bob can upper bound $\delta_p$ (denoted as
$\delta_p^u$) with this inequality given $\delta_b$. Clearly, when
$\delta_p=\delta_b$, the inequality will be always satisfied, i.e.,
$\delta_p=\delta_b$ is a particular solution of
Eq.~\eqref{Bounds:bp}.
Therefore, in general we have $\delta_p^u\ge\delta_p$. In the
following, we use $\delta_p$ to denote the upper bound $\delta_p^u$
for simplicity.



\begin{figure}[hbt]
\centering \resizebox{12cm}{!}{\includegraphics{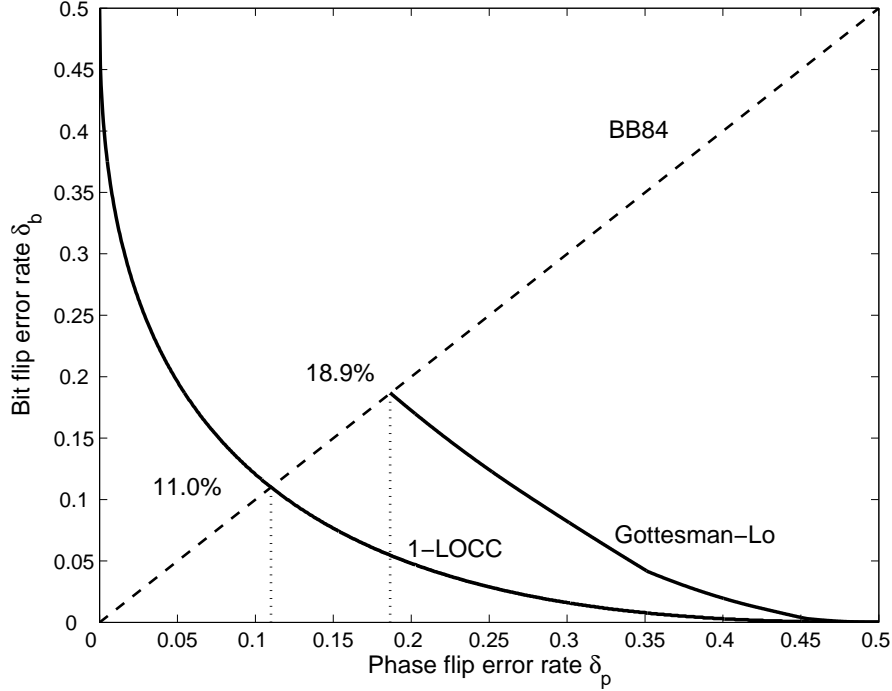}}
\caption{shows the secure regions in terms of error rates for 1-LOCC
EDP and Gottesman-Lo EDP. The regions under solid lines are proven
to be secure due to 1-LOCC EDP, and Gottesman-Lo EDP schemes (for
the region under the solid line and dashed line), respectively. For
1-LOCC EDP, we use Eq.~\eqref{EDP:CSSrate}. For Gottesman-Lo EDP, we
use Eqs.~\eqref{Twoway:oneBmatrix} and \eqref{Twoway:onePmatrix}. In
Gottesman-Lo EDP, we optimize the B/P sequence up to 12 steps. }
\label{Twoway:Fig:EDP}
\end{figure}

Given a QKD protocol and a laser source, Alice and Bob can estimate the phase error
rate $\delta_p$ from the bit error rate $\delta_b$ according to the protocol and the source. We investigate
the highest error rates that a data post-processing scheme can
tolerate.
Fig.~\ref{Twoway:Fig:EDP} shows
the tolerable error rates of the Gottesman-Lo EDP
compared to the 1-LOCC EDP scheme,
illustrating the superior performance of the Gottesman-Lo EDP
over the 1-LOCC EDP.
%
The boundaries of error rates are found by searching through the
regime of
\begin{equation} \label{Twoway:bpcondition}
\begin{aligned}
\delta_b&\le\delta_p \\
\delta_b+\delta_p&<1/2
\end{aligned}
\end{equation}
such that positive key rates are obtained. The reason we are
interested in the region specified the second inequality in
Eq.~\eqref{Twoway:bpcondition} is as follows: We can assume that the
error rates $\delta_b$ and $\delta_p$ are less than $1/2$, otherwise
Alice and Bob can flip the qubits. Also, if $\delta_b+\delta_p \ge
1/2$, then all the diagonal elements of the density matrix in
Eq.~\eqref{EDP:densitymatrix} Alice and Bob share are no greater
than $1/2$ (by setting $q_{11}=0$). Thus, the diagonalized density
matrix is separable \cite{BDSW} and the Gottesman-Lo EDP cannot distill any pure
EPR pairs.

The input to the Gottesman-Lo EDP is $(q_{00}, q_{10}, q_{11},
q_{01})$ with $q_{00}+q_{10}+q_{11}+q_{01}=1$, see in Subsection
\ref{BP}. But Alice and Bob only know $\delta_b=q_{10}+q_{11}$ and
$\delta_p=q_{11}+q_{01}$ from their error test. There is one free
parameter $q_{11}$. In Appendix C of \cite{GL}, the authors have
proved that $q_{11}=0$ is the worst case when $\delta_b=\delta_p$.
Following that proof, we can show that $q_{11}=0$ is the worst case
when the condition of Eq.~\eqref{Twoway:bpcondition} is satisfied.
That is, given $(\delta_b, \delta_p)$, if we check the input
$(1-\delta_b-\delta_p, \delta_b, 0, \delta_p)$ for Gottesman-Lo EDP
and get a positive key rate, then we can safely claim that
Gottesman-Lo EDP can tolerate the error rates of $(\delta_b,
\delta_p)$.

To determine the tolerable bit error rate of a particular protocol,
one should first obtain the relationship between the bit error rate
and the phase error rate, and plot it on FIG.~\ref{Twoway:Fig:EDP}.
The intersections between this curve and the boundary curves (the
1-LOCC curve and the Gottesman-Lo curve) indicate the tolerable QBER
for the corresponding EDPs. For example, for the BB84 protocol with
a perfect single-photon source, we have $\delta_b=\delta_p$, which
is the dashed line plotted in FIG.~\ref{Twoway:Fig:EDP}. We can
immediately read off from the figure that an initial bit error rate
of $18.9\%$ is tolerable using the Gottesman-Lo EDP \cite{GL}, while
a error rate of $11.0\%$ is tolerable using the 1-LOCC EDP. In
general, the Gottesman-Lo EDP gives rise to higher tolerable error
rates than the 1-LOCC EDP does.

For protocols having constraints on $q_{11}$, such as the six-state
protocol \cite{sixstate} and the SARG04 protocol with a
single-photon source \cite{SARG04,Tamaki,Fung}, the tolerable QBER
can go beyond the boundary curves shown in
FIG.~\ref{Twoway:Fig:EDP}.

We wrote a computer program
to exhaustively search for the optimal B/P sequence up to 12 steps.
The precision of the program is $10^{-15}$.



\subsection{Distance upper bound} \label{DisUp}
Let us come back to the realistic QKD setup model discussed in
Subsection \ref{Model}. An upper bound on the bit error rate of the
single photon state is 25\%, above which BB84 is broken by the
intercept-resend attack. The maximal secure distance then can be
bounded by the distance when the bit error rate of the single photon
state reaches 25\%.

The error rate of the single photon state $e_1$ is given in
\eqref{Model:ei},
$$
e_1=\frac{e_{d}\eta+\frac12Y_0}{\eta+Y_0}
$$
where $e_{d}$ is the intrinsic error rate of the detector in Bob's
side, $\eta$ is the overall transmittance, and $Y_0$ is the
background rate. Thus, $e_1$ exceeds 25\% when
\begin{equation} \label{Model:EtaBound}
\eta\le\frac{0.25Y_0}{0.25-e_{d}}.
\end{equation}
In GYS \cite{GYS}'s case, the fiber loss is $\alpha=0.21dB/km$,
$e_{d}=3.3\%$ and $Y_0=1.7\times10^{-6}$, then the upper bound of
secure distance is $208km$.

\subsection{Key generation rate upper bound} \label{KeyUp}
According to our model, the final secure key can only be derived
from single photon qubits. To derive the upper bound of key
generation rate, we assume that Alice and Bob
can distinguish the single photon qubits from other qubits (say,
vacuum and multi photon qubits). So they can perform the classical
data post-processing only on to the single photon qubits. One upper
bound of key generation rate is given by the \emph{mutual
information} between Alice and Bob \cite{Maurer1999},
\begin{equation}\label{Model:UpperR}
R^U=Q_1[1-H_2(e_1)],
\end{equation}
where $Q_1$ is the yields of single photon states and $e_1$ is the
error rate of single photon states. 

Note that the above two upper bounds, Eqs.~\eqref{Model:EtaBound}
and \eqref{Model:UpperR}, assume that a) Alice and Bob cannot
distinguish background counts and true signal counts and b) secure
key can only be extracted from the single photon states.
Also, these two bounds are general upper bounds regardless of the technique used for combating the effect of imperfect devices such as the decoy-state technique.

\section{Decoy + GLLP + Gottesman-Lo EDP} \label{BDecoy}
In this section, we propose a new 2-LOCC based data post-processing protocol with
a form of a sequence of B steps, followed by error correction and
privacy amplification, as discussed in Subsection \ref{BP}. This new
scheme is a generalization of the Gottesman-Lo scheme to the case of
imperfect devices. The reasons why we skip P steps here are as
follows. First, from the simulation in Subsection \ref{Boundary}, we
found that P steps are not as useful as B steps. Secondly, only
considering B steps can simplify the procedure of the data
post-processing scheme.

The procedure of this data post-processing is as follows.
\begin{enumerate}
\item
Alice and Bob perform a sequence of B steps to the sifted keys
(corresponding to $\tilde{r}_{B}$ in Eq.~\eqref{BDecoy:DGBrate}).
\item
They calculate the
variables (such as QBER, untagged qubits ratio) after the B steps.
\item They perform overall error correction (corresponding to the first
term in Eq.~\eqref{BDecoy:DGBrate}).
\item
They
perform
privacy amplification (corresponding to the second
term in Eq.~\eqref{BDecoy:DGBrate}).
\end{enumerate}
In the following, we will
discuss how to calculate the residue of this data post-processing
scheme.

In the decoy protocol, there are three kind of qubits: vacuum,
single photon and multi photon qubits, described in Section
\ref{Model}. We emphasize again here that the final secure key can
only be distilled from untagged qubits (single photon qubits).

Since either of the two inputs of a B step has three possibilities,
the outcomes of a B step then have nine possibilities. Only the case
that both inputs are untagged qubits has positive contribution to
the final secure key and all other privacy amplification terms in
Eq.~\eqref{GLLP:RateGeneral} will be $0$. That is, at the end of
some B steps and bit error correction, privacy amplification can be
only applied to the remaining qubits that come from the case where
both inputs are untagged qubits. In other words, an output qubit
after a subsequence of B steps is ``untagged" iff a) it passes all B
steps and b) it is generated from the case where all initial input
qubits are single photon qubits. Therefore, the residue ratio of
data post processing can be expressed, according to
Eq.~\eqref{GLLP:RateGeneral}, as
\begin{equation}\label{BDecoy:DGBrate}
\begin{aligned}
r=\max\{\tilde{r}_{B}[-f(\tilde{\delta})H_2(\tilde{\delta})+\tilde{\Omega}(1-H_2(\tilde{\delta}_p^{untagged}))],0\}
\end{aligned}
\end{equation}
where $\tilde{\delta}$ is the overall QBER, $\tilde{r}_{B}$ is
overall survival residue, $\tilde{\Omega}$ is the fraction of
untagged states in the final survival states and
$\tilde{\delta}_p^{untagged}$ is the phase error rate of the
untagged states, after a sequence of B steps. In the following, we
will show how these variables change with performing B steps.

\textbf{An arbitrary B step:} B step is an important two-way
primitive that we will use in this paper. Let us consider how the
various quantities (fraction of untagged states $\Omega$, QBER of
overall surviving states $\delta$, bit error rate
$\delta_{untagged}$ and phase error rates $\delta_p$ of the untagged
states) are transformed under one step in a B step sequence.

Before a B step, the fraction of untagged states is $\Omega$, the
overall QBER is $\delta$, the bit error rate of the untagged states
is $\delta_{untagged}$, and the phase error rate of the untagged
states is $\delta_p$. According to Eq.~\eqref{oneBpS} the overall
survival probability $p_{S}$ and the survival probability of the
untagged states $p_{S}^{untagged}$ after one B step are given by
\begin{equation} \label{BDecoy:1BpS}
\begin{aligned}
p_{S}&=[\delta^2+(1-\delta)^2]\\
p_{S}^{untagged}&=[\delta_{untagged}^2+(1-\delta_{untagged})^2].
\end{aligned}
\end{equation}
Then the residue after one B step is given by,
\begin{equation} \label{B:1Bresidue}
\begin{aligned}
r_{B}=\frac12p_S
\end{aligned}
\end{equation}
The factor $\frac12$ in Eq.~\eqref{B:1Bresidue} due to the fact that
Alice and Bob only keep one qubit from a survival pair. Then, after
a B step the fraction of untagged states $\Omega'$ is given by
\begin{equation} \label{BDecoy:1Bfrac}
\begin{aligned}
\Omega'&=\frac{\Omega^2\cdot p_{S}^{untagged}}{p_{S}}.
\end{aligned}
\end{equation}

Overall QBER: the change of overall QBER $\delta'$ is given by
\begin{equation} \label{BDecoy:1BQBER}
\begin{aligned}
\delta' &= \frac{\delta^2}{\delta^2+(1-\delta)^2}.
\end{aligned}
\end{equation}

Untagged states: before the first round of B step, the initial
density matrix of untagged state is
$(1-2e_1+q_{11},e_1-q_{11},q_{11},e_1-q_{11})$, where $e_1$ is the
error rate of single photon states. From Appendix C of \cite{GL},
we know that $q_{11}=0$ is the worst case for B steps. Thus we can
conservatively choose $(1-2e_1,e_1,0,e_1)$ as the initial input
density matrix. If only B steps are performed, $q_{11}=0$ will
always be satisfied, according to Eq.~\eqref{Twoway:oneBmatrix}. So
the input untagged qubits for any round of B step has the form of
\begin{equation} \label{BDecoy:1Binput}
(q_{00}, q_{10}, q_{11}, q_{01}) = (1-\delta_{untagged}-\delta_p,
\delta_{untagged}, 0, \delta_p).
\end{equation}
The bit error rate of untagged state $\delta_{untagged}'$ only
depends on the input $\delta_{untagged}$,
\begin{equation} \label{BDecoy:1Bbun}
\delta_{untagged}'=\frac{\delta_{untagged}^2}{\delta_{untagged}^2+(1-\delta_{untagged})^2}.
\end{equation}
According to Eqs.~\eqref{Twoway:oneBmatrix},
\eqref{Twoway:AfterBstepErr} and \eqref{BDecoy:1Binput}, the phase
error rate of untagged states is
\begin{equation} \label{BDecoy:1Bpun}
\begin{aligned}
\delta_p'&=q_{11}'+q_{01}'\\
&=\frac{2q_{10}q_{11}+2q_{00}q_{01}}{(q_{10}+q_{11})^2+(q_{00}+q_{01})^2}\\
&=\frac{2\delta_p\cdot(1-\delta_{untagged}-\delta_p)}{\delta_{untagged}^2+(1-\delta_{untagged})^2}.
\end{aligned}
\end{equation}

Eqs.~\eqref{BDecoy:1BpS}-\eqref{BDecoy:1Bpun} are valid for a
general B step. Alice and Bob can perform a sequence of B steps as
described above and then do the error correction and privacy
amplification. Once all the these quantities are obtained, the key
generation rate can be calculated from Eq.~\eqref{BDecoy:DGBrate}.

To illustrate the improvement made by introducing B steps, we
numerically calculated the key generation rate assuming the
parameters in the GYS experiment
\cite{GYS}. Note that the overall QBER $\tilde{\delta}$ in
Eq.~\eqref{BDecoy:DGBrate} never exceeds 10\%. The value of
$f(e)=1.22$ is the upper bound according to \cite{BS}. The
parameters used for simulation are listed in Table
\ref{Tab:GYSdata}.

\begin{table}[hbt]
\centering
\begin{tabular}{|c|c|c|c|c|} \hline
Wavelength [nm] & $\alpha$ [dB/km] & $\eta_{Bob}$ & $e_{d}$ & $Y_0$ \\
\hline
1550 & 0.21 & 4.5\% & 3.3\% & $1.7\times 10^{-6}$ \\
\hline
\end{tabular}
\caption{Data come from GYS \cite{GYS}.} \label{Tab:GYSdata}
\end{table}

\begin{figure}[hbt]
\centering \resizebox{12cm}{!}{\includegraphics{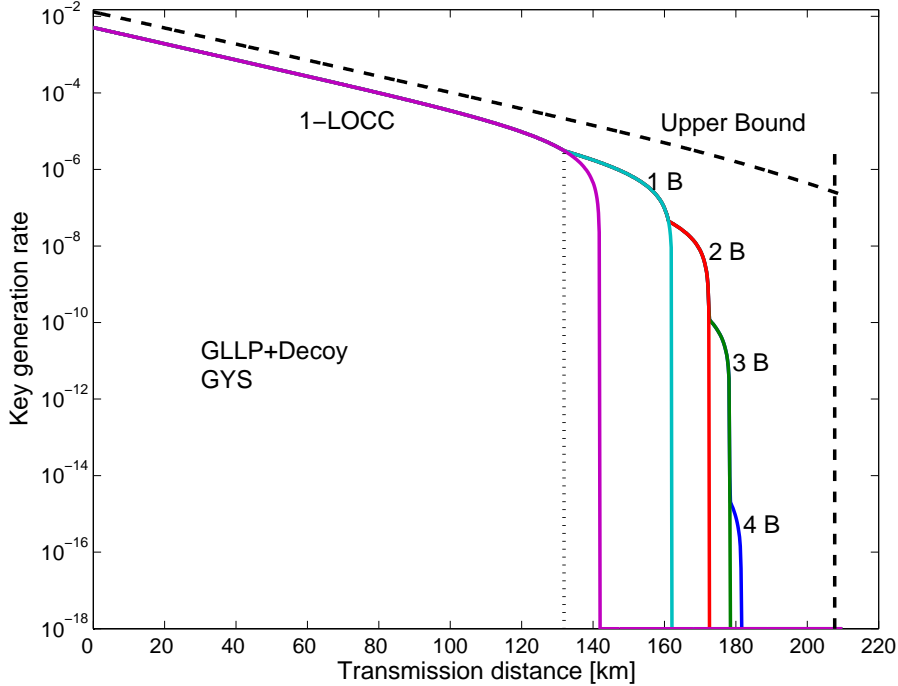}}
\caption{shows the key generation rate as a function of the
transmission distance with the data post-processing scheme of the
GLLP+Decoy+B steps. The parameters used are from the GYS experiment \cite{GYS} listed in Table
\ref{Tab:GYSdata}. GLLP+Decoy+B steps scheme suppresses the one with
1-LOCC at distance of 132km.
The maximal secure distance using 4 B steps is $181km$, which is not
far from the upper bound of $208km$.
}
\label{Fig:5B}
\end{figure}

From FIG.~\ref{Fig:5B}, we can see that there is a non-trivial
extension of maximal secure distance after introducing B steps. We
remark that the key generation rate decoy state protocol with 1 B
step is higher than the one with 1-LOCC from the distance around
$132km$. The maximal secure distance using 4 B steps is $181km$,
which is not far from the upper bound of $208km$, given in Section
\ref{DisUp}. Even with only one B step, the maximal secure distance
can be extended from 142km to 162km. Thus, B steps are very useful
in QKD data post-processing.

\section{Decoy + GLLP + Recurrence EDP} \label{Recurrence}
In this section, we propose a second 2-LOCC based data post-processing scheme
based on the recurrence scheme \cite{GVV}, which is reviewed in
Subsection~\ref{Rec:Subsection:Rec}. Our scheme is a generalization
of the recurrence scheme to the case of imperfect sources.

In Section \ref{GLLP}, we give out a formula,
Eq.~\eqref{GLLP:RateGeneral}, for key generation rate with the idea
of GLLP. Let us combine recurrence with
Eq.~\eqref{GLLP:RateGeneral}. So, instead of just taking care of one
kind of qubit, we need to apply privacy amplification to several
groups of qubits separately, i.e., we will have several $K_i$ in
Eq.~\eqref{Rec:Residue0}. After the recurrence, the data
post-processing residue rate becomes
\begin{equation} \label{Rec:Residue1}
\begin{aligned}
r=-\frac12f(p_S)H_2(p_S)-\frac12p_Sf(\frac{\delta^2}{p_S})H_2(\frac{\delta^2}{p_S})+\sum_{i}\Omega_iK_{i},
\end{aligned}
\end{equation}
where $p_S$ is the even parity possibility given in
Eq.~\eqref{Rec:Survival} with $\delta_b^C=\delta_b^T=\delta$,
$\delta$ is the overall QBER before the recurrence, $f(\cdot)$ is
error correction efficiency, $\Omega_i$ and $K_{i}$ are the
probability and the residue of the qubit groups with label $i$ after
privacy amplification, respectively. Here, Alice and Bob first check
the parity, corresponding to the first term of
Eq.~\eqref{Rec:Residue1}. Secondly, they apply overall error
correction to the qubits with even parity, corresponding to the
second term of Eq.~\eqref{Rec:Residue1}. Thirdly, they measure one
of qubits in those pairs with odd parity to obtain the error
syndrome of another qubit. Afterwards, they can group the surviving
qubits into several groups with labels $i$. Finally, they perform
privacy amplification to each group with label $i$, corresponding to
the last term of Eq.~\eqref{Rec:Residue1}.

Consider the decoy state case, Alice and Bob have three kinds of
input qubits: vacuum qubits (V), single photons qubit (S) and multi
photon qubits (M). The input parameters for recurrence are listed in
Table \ref{Rec:Tab:Input}.
\begin{table}[hbt]
\centering
\begin{tabular}{|c|c|c|c|c|}
\hline
Qubit & Fraction & $\delta_b$ & $\delta_p$ & $q_{11}$ \\
\hline
V & $\Omega_V$ & $1/2$ & $1/2$ & $q_{11}^V$ \\
\hline
S & $\Omega$ & $e_1$ & $e_1$ & $a$ \\
\hline
M & $\Omega_M$ & $e_M$ & $1/2$ & $q_{11}^M$ \\
\hline
\end{tabular}
\caption{lists the input parameters of three kinds of qubits for
recurrence. Following Eq.~\eqref{Model:Qi} and \eqref{Model:Gain},
the fractions of each group are given by $\Omega_V=Q_0/Q_\mu$,
$\Omega=Q_1/Q_\mu$ and $\Omega_M=1-\Omega_V-\Omega$.
$\Omega_V/2+e_1\Omega+e_M\Omega_M=\delta$ is the overall QBER.}
\label{Rec:Tab:Input}
\end{table}

Thus, the outcome of one round of recurrence will have nine cases.
Clearly, if neither input is a single photon qubits, the outcome
will have no contribution to the final key. Alice and Bob need only
apply Eq.~\eqref{Rec:PriRes2} to calculate the residues, $K_i$, for
the five cases: $V \bigotimes S$, $S \bigotimes V$, $S\bigotimes S$,
$S\bigotimes M$, $M\bigotimes S$. The probabilities of occurrence,
$\Omega_i$, for the five cases are, respectively, $\Omega_V\Omega$,
$\Omega\Omega_V$, $\Omega^2$, $\Omega\Omega_M$, $\Omega_M\Omega$.
Once we know $K_i$ and $\Omega_i$, we can then determine the overall
residue, $r$, using Eq.~\eqref{Rec:Residue1} (details are shown in
Appendix~\ref{App:residue}):
\begin{equation} \label{Rec:Residue2}
\begin{aligned}
r 
\ge& -B+C-F_a
\end{aligned}
\end{equation}
where
\begin{equation}
\begin{aligned}
B &=
\frac12f(p_S)H_2(p_S)+\frac12p_Sf(\frac{\delta^2}{p_S})H_2(\frac{\delta^2}{p_S})
\\
C &= \frac34\Omega_V\Omega + \Omega^2 (1-e_1+e_1^2) +
\frac12\Omega\Omega_M(2-e_1-e_M+2e_1e_M)
\\
D_1 &=
\frac34\Omega_V\Omega+\frac12\Omega^2(2-e_1)+\frac12\Omega\Omega_M(2-e_M)
\\
D_2 &=
\frac34\Omega_V\Omega+\frac12\Omega^2(1+e_1)+\frac12\Omega\Omega_M(e_M+1)
\\
F_a&=D_1(1-e_1)H_2(\frac{e_1-a}{1-e_1})+D_2e_1H_2(\frac{a}{e_1}) \\
\end{aligned}
\end{equation}
with equality when $q_{11}^V=1/4$ and $q_{11}^M=e_M/2$. In order to
get a lower bound of key generation rate $R$, we maximize $F_a$ over
$a$ by using a bisection method as discussed in Appendix
\ref{App:residue}.


\begin{figure}[hbt]
\centering \resizebox{12cm}{!}{\includegraphics{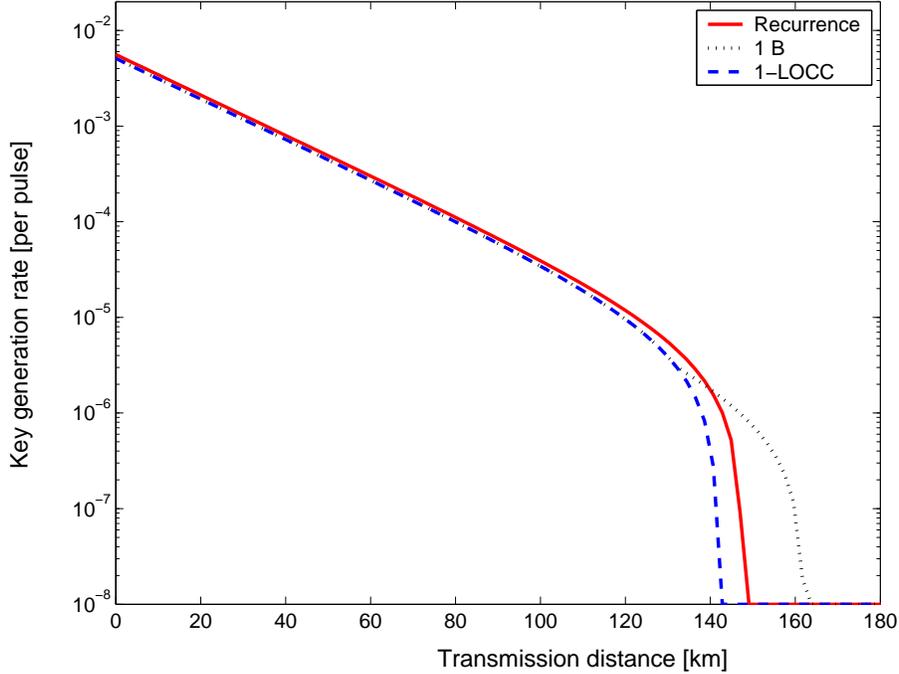}}
\caption{Plot of the key generation rate as a function of the
transmission distance, GLLP+Decoy+Recurrence vs. GLLP+Decoy+1-LOCC.
Recurrence does have some marginal improvement over 1-LOCC for short
distances.
In particular, the recurrence method increases the key generation
rate by more than $10\%$ in our simulation. The maximal secure
distance for each case is 142.8km (1-LOCC), 149.1km (Recurrence),
163.8km (1B), respectively. Here we consider the asymptotic Decoy
state QKD case with infinitely long signals.
The parameters used are from the GYS experiment \cite{GYS} listed in Table
\ref{Tab:GYSdata}.} \label{Rec:Fig:Comp}
\end{figure}

Figure \ref{Rec:Fig:Comp} shows the key generation rate as a
function of the transmission distance for GLLP+Decoy+1-LOCC,
GLLP+Decoy+1B, and GLLP+Decoy+Recurrence. Recurrence does have some
marginal improvement (more than 10\%) in the key generation rate
over 1-LOCC for short distances, and it also increases the maximal
secure distance by $6km$. We remark that recurrence is useful even
in the short distance regime.

\section{Statistical fluctuations} \label{StaFlu}
In a realistic QKD experiment, only a finite number of signals are transmitted.
Thus, the estimations of $Y_1$ and $e_1$ have certain statistical fluctuations.
These statistical fluctuations in decoy state QKD with 1-LOCC are analyzed
in \cite{Practical}, and also \cite{WangDecoy}.
In this section, we will consider statistical fluctuations for two
data post-processing schemes with 2-LOCC discussed in Section
\ref{BDecoy} and Section \ref{Recurrence}.
Following the analysis in \cite{Practical}, we can easily incorporate statistical
fluctuations in these two data post-processing schemes.


In \cite{Practical}, $Q_1$ and $e_1$ are bounded when taking
statistical fluctuations into account. As for the post-processing
with 2-LOCC described in Section \ref{BDecoy}, the inputs of the B
steps are $Q_\mu$, $E_\mu$, which can be measured directly from
experiment, and $Q_1$, $e_1$, which are estimated by the decoy
method. Evidently, one should take the lower bound of $Q_1$ and the
upper bound of $e_1$ to lower bound the key rate $R$. Thus the
procedure will be as follows, first with the decoy method, one can
lower bound $Q_1$ and upper bound $e_1$, and then input the four
parameters ($Q_\mu$, $E_\mu$, $Q_1$ and $e_1$) into the data
post-processing of GLLP+Decoy+B step to extract secure keys.

As for the case of recurrence, from Table \ref{Rec:Tab:Input}, Alice
and Bob have to estimate $Q_0$ besides the four parameters discussed
above. From Eq.~\eqref{Rec:Residue2}, it is not clear which bounds
of $Q_0$ one should pick up to lower-bound the key rate. One can
clearly see that lower-bounding Eq.~\eqref{Rec:Residue2} is a hard
problem. Instead of going to tedious mathematical calculations here,
we have a plausible argument based on our physical intuition.
First of all, single photon qubits are ``good" qubits in our
discussion. So, we reasonably assume that Alice and Bob can safely
use the lower bound of $Q_1$ and the upper bound of $e_1$ to lower
bound the key rate. As discussed in \cite{nothing}, the vacuum decoy
state can have positive contribution in the privacy amplification
procedure. Also, in the Appendix of \cite{Practical}, we have proven
that one should take the lower bound of $Y_0$ to lower bound the key
generation rate in 1-LOCC case. Thus, here we take the lower bound
of $Q_0$ to estimate the key rate.

\begin{figure}[hbt]
\centering \resizebox{12cm}{!}{\includegraphics{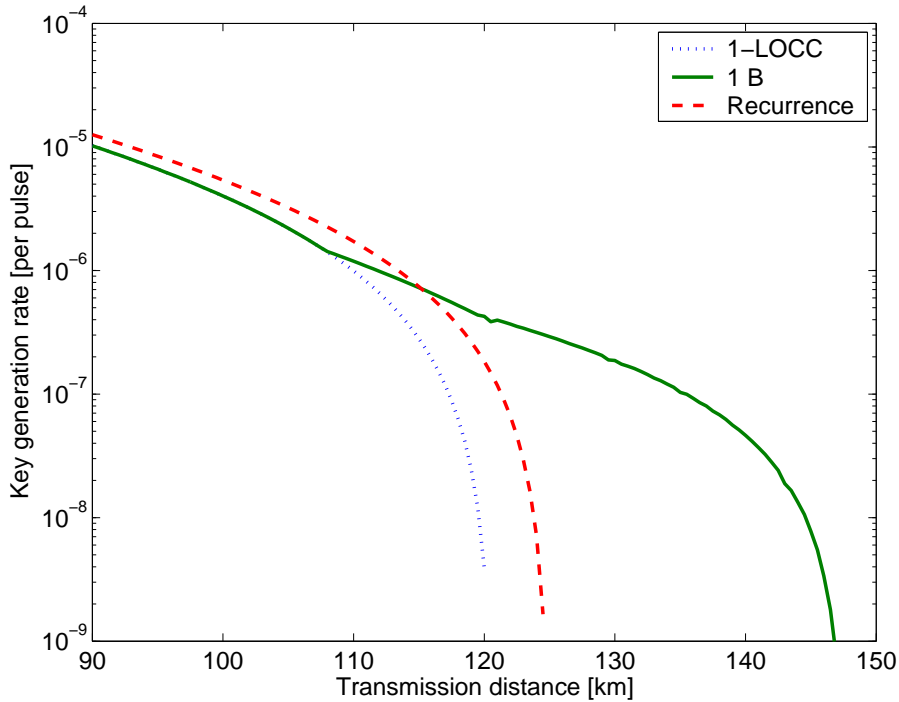}}
\caption{shows the simulation results for three data post-processing
schemes of the decoy state protocol, Decoy+1-LOCC, Decoy+1 B and
Decoy+Recurrence, considering statistical fluctuations. The maximal
secure distances of three schemes are 120km, 125km and 147km,
respectively.
The parameters used are from the GYS experiment \cite{GYS} listed in Table
\ref{Tab:GYSdata}.} \label{Rec:Fig:StaFlu}
\end{figure}

The results are shown in Figure \ref{Rec:Fig:StaFlu}.
Here, we assume Alice and Bob use $6 \times 10^{9}$ number of pulses.
These pulses are randomly selected as signal and decoy states.
The distribution, among the signal states, vacuum states and weak decoy states, was found by an exhaustive search for the optimal one.
Then,
we use
10 standard deviations to bound $Q_\nu$, $E_\nu$ and $Y_0$, and substitute the worst-case values of these into
\begin{equation}\label{Decoy:VWQ1Bound}
\begin{aligned}
Q_1 \ge Q_1^{L,\nu,0} &= \frac{\mu^2e^{-\mu}}{\mu\nu-\nu^2}(Q_\nu
e^{\nu}-Q_\mu
e^\mu\frac{\nu^2}{\mu^2}-\frac{\mu^2-\nu^2}{\mu^2}Y_0) \\
e_1 \le e_1^{U,\nu,0} &= \frac{E_\nu Q_\nu
e^{\nu}-e_0Y_0}{Y_1^{L,\nu,0}\nu}
\end{aligned}
\end{equation}
to bound $Q_1$ and $e_1$.
Here, $\nu$ is the expected photon number of the weak decoy states, $Q_\nu$ and $E_\nu$ are the gain and QBER of the weak decoy states.
We can see
that with statistical fluctuations, the improvements of 2-LOCC are
still notable.

\section{Conclusion}
We have developed two data post-processing schemes for decoy-state
QKD using 2-LOCC, one based on B steps and the other one based on the
recurrence method.
The distance of secure QKD is crucial in practical applications.
Therefore, 
our Decoy+B steps
post-processing protocol, which we have shown to be able to increase the maximal secure distance of
QKD from about 141 km to about 182 km (using parameters from the GYS experiment \cite{GYS}), proves useful in
real-life applications. Moreover, our work shows that recurrence
protocols are useful for increasing the key generation rate in a
practical QKD system even at short distances. While we have focused
our modeling on a fiber-based QKD system, our general formalism
applies also to open-air QKD systems.

We have shown that similar conclusions hold even with statistical
fluctuations in the experimental variables for the Decoy+B step
scheme. For the Decoy+Recurrence scheme, although we do not have a
rigorous argument, physical intuition suggests that similar
conclusions hold with statistical fluctuations as well. We conclude
that using two-way classical communications is superior to using
one-way for our decoy-state QKD schemes.

In addition, we provided the region of bit error rates and phase
error rates that are tolerable by using the Gottesman-Lo EDP scheme.
Also, we calculated the upper bounds on distance and on the key
generation rate of a real QKD setup based on our model.


%

 Acknowledgments
\section{Acknowledgments}
We thank G.~Brassard, B.~Fortescue, D.~Gottesman, and B.Qi for
enlightening discussions. Financial support from CFI, CIAR, CIPI,
Connaught, CRC, NSERC, OIT, PREA and the University of Toronto is
gratefully acknowledged.

\begin{appendix}

\section{Key rate of the Recurrence scheme with an ideal source\label{App:review-recurrence}}

In this section, we review the recurrence EDP and develop the key generation rate formula given by
\begin{equation}
R=q \cdot r,
\end{equation}
where
$q$ depends on the implementation of the QKD ($1/2$ for the BB84
protocol, because half the time Alice and Bob bases are not
compatible) and
$r$ is the residue which we will find in the sequel.
In the following,
we use the same notation as in Subsection \ref{review-EDP} and consider a
Bell diagonal state $(q_{00}, q_{10}, q_{11}, q_{01})$.

\subsubsection{Parity check}  As the first
step of recurrence, Alice and Bob check the parity of two pairs
(labeled by control qubit $C$ and target qubit $T$). They will get
even parity if the two pairs are in one of the states
$$
0000,0001,0100,0101,1010,1011,1110,1111,
$$
and will get odd parity if they are in one of the states
$$
0010,0011,0110,0111,1000,1001,1100,1101,
$$
where the first two bits represent the control qubit, and the last
two bits represent the target qubit. For example, $1110$ means that
there is a bit error and a phase error in the control qubit, and a
bit error and no phase error in the target qubit. Thus, the
probability to get even parity is given by
\begin{equation} \label{Rec:Survival}
\begin{aligned}
p_S &= (q_{00}^C+q_{01}^C)(q_{00}^T+q_{01}^T) + (q_{10}^C+q_{11}^C)(q_{10}^T+q_{11}^T)\\
    &= (1-\delta_b^C)(1-\delta_b^T) + \delta_b^C\delta_b^T,
\end{aligned}
\end{equation}
where $\delta_b^C=q_{10}^C+q_{11}^C$ and
$\delta_b^T=q_{10}^T+q_{11}^T$ are the bit error rates of the input
control and target qubits, respectively. During the parity check,
the number of pure EPR pairs that Alice and Bob need to sacrifice is
given by
\begin{equation} \label{Rec:ParityCh}
\begin{aligned}
\frac12H_2(p_S),
\end{aligned}
\end{equation}
where $\frac12$ is due to the fact that Alice and Bob compute the
parity of two-qubit pairs at one time.
After the parity check, the qubits are divided into two groups,
qubits with even parity and odd parity. In the following, we will
discuss the error correction and privacy amplification on these two
groups separately. The recurrence protocol appearing in \cite{GVV}
only performs error correction on qubits with even parity.

\subsubsection{Error correction}
For even parity qubits, we can see that the bit error syndrome of
control qubits will be the same as that of target qubits. Thus,
Alice and Bob only need to do error correction on the control (or
target) qubits. According to Eq.~\eqref{Twoway:AfterBstepErr}, the
bit error rate of control qubits after recurrence is given by
\begin{equation} \label{Rec:ErrControl}
\begin{aligned}
\tilde\delta_b^{C} =
\frac{(q_{10}^C+q_{11}^C)(q_{10}^T+q_{11}^T)}{p_S} =
\frac{\delta_b^C\delta_b^T}{p_S}
\end{aligned}
\end{equation}
where $p_S$ is the probability of even parity in the recurrence
given by Eq.~\eqref{Rec:Survival}. Therefore, Alice and Bob need to
sacrifice a fraction
\begin{equation} \label{Rec:OverallErrC}
\begin{aligned}
\frac12p_SH_2(\tilde\delta_b^{C}) =
\frac12p_SH_2(\frac{\delta_b^C\delta_b^T}{p_S})
\end{aligned}
\end{equation}
to do the overall error correction. The factor $\frac12$ is due to
the fact that control qubits have the same error syndrome as target
qubits.

Therefore the residue of data post-processing, similar to
Eq.~\eqref{GLLP:GLLPex}, can be expressed as
\begin{equation} \label{Rec:Residue0}
\begin{aligned}
r = -\frac12H_2(p_S)
-\frac12p_SH_2(\frac{\delta_b^C\delta_b^T}{p_S}) + K
\end{aligned}
\end{equation}
where $p_S$ is given in Eq.~\eqref{Rec:Survival} and K is the
residue of privacy amplification, which we will focus on in the
following.

\subsubsection{Privacy amplification}
Alice and Bob perform privacy amplification to the qubits with even
and odd parity separately.

\textbf{Even parity:} now, Alice and Bob already know the bit error
syndrome. The control and target qubits have the same bit error
syndromes, but may have different phase error syndromes. Thus, Alice
and Bob can divide the even parity qubits into four groups: control
qubits with bit error syndrome 0 and 1,  and target qubits with bit
error syndrome 0 and 1, and treat these groups separately in the
privacy amplification step. The probability of each group (summing
together the even parity probabilities given in
Eq.~\eqref{Rec:Survival}) is given by
$$
\frac{(q_{00}^C+q_{01}^C)(q_{00}^T+q_{01}^T)}{2},
\frac{(q_{10}^C+q_{11}^C)(q_{10}^T+q_{11}^T)}{2},
\frac{(q_{00}^C+q_{01}^C)(q_{00}^T+q_{01}^T)}{2},
\frac{(q_{10}^C+q_{11}^C)(q_{10}^T+q_{11}^T)}{2}
$$
with phase error rate
$$
\frac{q_{01}^C}{q_{00}^C+q_{01}^C},
\frac{q_{11}^C}{q_{10}^C+q_{11}^C},
\frac{q_{01}^T}{q_{00}^T+q_{01}^T},
\frac{q_{11}^T}{q_{10}^T+q_{11}^T}.
$$
Since the error syndrome of each group of qubits is known to Alice and Bob,
privacy amplification can be applied to the different
groups separately.
Then, Alice
and Bob should sacrifice a fraction
\begin{equation} \label{Rec:PriAmpPri}
\begin{aligned}
&\frac{(q_{00}^C+q_{01}^C)(q_{00}^T+q_{01}^T)}{2}H_2(\frac{q_{01}^C}{q_{00}^C+q_{01}^C})
+
\frac{(q_{10}^C+q_{11}^C)(q_{10}^T+q_{11}^T)}{2}H_2(\frac{q_{11}^C}{q_{10}^C+q_{11}^C}) + \\
&\frac{(q_{00}^C+q_{01}^C)(q_{00}^T+q_{01}^T)}{2}H_2(\frac{q_{01}^T}{q_{00}^T+q_{01}^T})
+
\frac{(q_{10}^C+q_{11}^C)(q_{10}^T+q_{11}^T)}{2}H_2(\frac{q_{11}^T}{q_{10}^T+q_{11}^T}) \\
\end{aligned}
\end{equation}
to do the privacy amplification. Given the bit and phase error rates
of input control and target qubits $\delta_p^C=q_{11}^C+q_{01}^C$
and $\delta_p^T=q_{11}^T+q_{01}^T$, Eq.~\eqref{Rec:PriAmpPri} can be
written as
\begin{equation} \label{Rec:PriAmp}
\begin{aligned}
\frac12(1-\delta_b^C)(1-\delta_b^T)[H_2(\frac{\delta_p^C-q_{11}^C}{1-\delta_b^C})+H_2(\frac{\delta_p^T-q_{11}^T}{1-\delta_b^T})] + \frac12\delta_b^C\delta_b^T[H_2(\frac{q_{11}^C}{\delta_b^C})+H_2(\frac{q_{11}^T}{\delta_b^T})]. \\
\end{aligned}
\end{equation}

Thus the privacy amplification residue of even parity qubits is
given by,
\begin{equation} \label{Rec:EvenRes}
\begin{aligned}
K_{even} =  p_S
-\frac12(1-\delta_b^C)(1-\delta_b^T)[H_2(\frac{\delta_p^C-q_{11}^C}{1-\delta_b^C})+H_2(\frac{\delta_p^T-q_{11}^T}{1-\delta_b^T})]
- \frac12\delta_b^C\delta_b^T[H_2(\frac{q_{11}^C}{\delta_b^C})+H_2(\frac{q_{11}^T}{\delta_b^T})]. \\
\end{aligned}
\end{equation}

\textbf{Odd parity:}
it turns out that pairs with odd parity during the recurrence can
also contribute to the final key \cite{GVV}. Instead of including
them in the error correction, Alice and Bob measure one of the two
qubits and hence they know the bit error syndrome of the remaining
qubit. They can then proceed with privacy amplification on those
qubits. 

Suppose Alice and Bob always choose to measure the target qubits and
obtain the error syndrome of the control qubits. Similar to the even
parity case, now, Alice and Bob can divide the control qubits with
odd parity into two parts according to the bit error syndrome. The
probability of each part is given by
$$
\frac{(q_{00}^C+q_{01}^C)(q_{10}^T+q_{11}^T)}{2},
\frac{(q_{10}^C+q_{11}^C)(q_{00}^T+q_{01}^T)}{2},
$$
with phase error rate
$$
\frac{q_{01}^C}{q_{00}^C+q_{01}^C},
\frac{q_{11}^C}{q_{10}^C+q_{11}^C}.
$$

With the same argument as Eq.~\eqref{Rec:PriAmpPri}, the number of
qubits that need be sacrificed to privacy amplification is given by
\begin{equation} \label{Rec:PriAmpPriOdd}
\begin{aligned}
&\frac{(q_{00}^C+q_{01}^C)(q_{10}^T+q_{11}^T)}{2}H_2(\frac{q_{01}^C}{q_{00}^C+q_{01}^C})
+
\frac{(q_{10}^C+q_{11}^C)(q_{00}^T+q_{01}^T)}{2}H_2(\frac{q_{11}^C}{q_{10}^C+q_{11}^C}) \\
&=\frac12[(1-\delta_b^C){\delta_b^T}H_2(\frac{\delta_p^C-q_{11}^C}{1-\delta_b^C})
+ \delta_b^C(1-\delta_b^T)H_2(\frac{q_{11}^C}{\delta_b^C})]
\\
\end{aligned}
\end{equation}

So the privacy amplification residue of odd parity qubits is given
by,
\begin{equation} \label{Rec:OddRes}
\begin{aligned}
K_{odd}=\frac12(1-\delta_b^C){\delta_b^T}[1-H_2(\frac{\delta_p^C-q_{11}^C}{1-\delta_b^C})]
+
\frac12\delta_b^C(1-\delta_b^T)[1-H_2(\frac{q_{11}^C}{\delta_b^C})]
\end{aligned}
\end{equation}

Therefore, the privacy amplification residue, $K$ in
Eq.~\eqref{Rec:Residue0}, by adding Eq.~\eqref{Rec:EvenRes} and
Eq.~\eqref{Rec:OddRes} and substituting Eq.~\eqref{Rec:Survival}, is
given by
\begin{equation} \label{Rec:PriRes2}
\begin{aligned}
K =& K_{even}+K_{odd} \\
=&1-\frac12(1-\delta_b^C){\delta_b^T}-\frac12\delta_b^C(1-\delta_b^T)
-\frac12(1-\delta_b^C)H_2(\frac{\delta_p^C-q_{11}^C}{1-\delta_b^C})
-\frac12\delta_b^CH_2(\frac{q_{11}^C}{\delta_b^C})
\\
&-\frac12(1-\delta_b^C)(1-\delta_b^T)H_2(\frac{\delta_p^T-q_{11}^T}{1-\delta_b^T})
-\frac12\delta_b^C\delta_b^TH_2(\frac{q_{11}^T}{\delta_b^T}).
\end{aligned}
\end{equation}
Note that there are two free parameters $q_{11}^C$ and $q_{11}^T$ in
Eq.~\eqref{Rec:PriRes2}, which should be minimized over to obtain the worst-case key rate.

\section{Residue for the Decoy+GLLP+Recurrence scheme\label{App:residue}}
We calculate the residues, $K_i$, in Eq.~\eqref{Rec:Residue1} for
the five cases: $V \bigotimes S$, $S \bigotimes V$, $S\bigotimes S$,
$S\bigotimes M$, $M\bigotimes S$. Here, we apply each case, with
parameters shown in Table \ref{Rec:Tab:Input} into
Eq.~\eqref{Rec:PriRes2} to calculate each $K_i$.

\textbf{$V \bigotimes S$:}  the probability of this case is
$\Omega_{VS}=\Omega_V\Omega$.
\begin{equation} \label{Rec:KVS}
\begin{aligned}
K_{VS} &= 1-\frac14-\frac14H_2(1-2q_{11}^V)-\frac14H_2(2q_{11}^V)
-\frac14(1-e_1)H_2\left(\frac{e_1-a}{1-e_1}\right) -
\frac14e_1H_2\left(\frac{a}{e_1}\right) \\
&\ge \frac14 -\frac14(1-e_1)H_2\left(\frac{e_1-a}{1-e_1}\right)
-\frac14e_1H_2\left(\frac{a}{e_1}\right)
\end{aligned}
\end{equation}
with equality when $q_{11}^V=1/4$. This is due to the concavity of
function $H_2(\cdot)$.

\textbf{$S \bigotimes V$:} the probability of this case is
$\Omega_{VS}=\Omega_V\Omega$.
\begin{equation} \label{Rec:KSV}
\begin{aligned}
K_{SV} &\ge
1-\frac14-\frac12(1-e_1)H_2\left(\frac{e_1-a}{1-e_1}\right)-\frac12e_1H_2\left(\frac{a}{e_1}\right)
-\frac14(1-e_1)H_2\left(1-2q_{11}^V\right) -
\frac14e_1H_2\left(2q_{11}^V\right) \\
&\ge
\frac12-\frac12(1-e_1)H_2\left(\frac{e_1-a}{1-e_1}\right)-\frac12e_1H_2\left(\frac{a}{e_1}\right)
\end{aligned}
\end{equation}
with equality when $q_{11}^V=1/4$.

\textbf{$S \bigotimes S$:} the probability of this case is
$\Omega_{VV}=\Omega^2$.
\begin{equation} \label{Rec:KSS}
\begin{aligned}
K_{SS} = 1 &- e_1(1-e_1) -
\frac12(1-e_1)H_2\left(\frac{e_1-a}{1-e_1}\right) -
\frac12e_1H_2\left(\frac{a}{e_1}\right) \\
&-\frac12(1-e_1)^2H_2\left(\frac{e_1-a}{1-e_1}\right) -
\frac12e_1^2H_2\left(\frac{a}{e_1}\right). \\
\end{aligned}
\end{equation}

\textbf{$S \bigotimes M$:} the probability of this case is
$\Omega_{SM}=\Omega\Omega_M$
\begin{equation} \label{Rec:KSM}
\begin{aligned}
K_{SM} &=
1-\frac12e_1(1-e_M)-\frac12e_M(1-e_1) -\frac12(1-e_1)H_2\left(\frac{e_1-a}{1-e_1}\right)-\frac12e_1H_2\left(\frac{a}{e_1}\right)\\
& -\frac12(1-e_1)(1-e_M)H_2\left(\frac{1-2q_{11}^M}{2-2e_M}\right)-\frac12e_1e_MH_2\left(\frac{q_{11}^M}{e_M}\right) \\
&\ge\frac12-\frac12(1-e_1)H_2\left(\frac{e_1-a}{1-e_1}\right)-\frac12e_1H_2\left(\frac{a}{e_1}\right),
\end{aligned}
\end{equation}
with equality when $q_{11}^M=e_M/2$.

\textbf{$M \bigotimes S$:} the probability of this case is
$\Omega_{MS}=\Omega_M\Omega$
\begin{equation} \label{Rec:KMS}
\begin{aligned}
K_{MS} &=
1-\frac12e_M(1-e_1)-\frac12e_1(1-e_M) -\frac12(1-e_M)H_2\left(\frac{1-2q_{11}^M}{2-2e_M}\right)-\frac12e_MH_2\left(\frac{q_{11}^M}{e_M}\right)\\
& -\frac12(1-e_1)(1-e_M)H_2\left(\frac{e_1-a}{1-e_1}\right)-\frac12e_1e_MH_2\left(\frac{a}{e_1}\right) \\
&\ge\frac12-\frac12e_M(1-e_1)-\frac12e_1(1-e_M)\\
& -\frac12(1-e_1)(1-e_M)H_2\left(\frac{e_1-a}{1-e_1}\right)-\frac12e_1e_MH_2\left(\frac{a}{e_1}\right), \\
\end{aligned}
\end{equation}
with equality when $q_{11}^M=e_M/2$.

Therefore, after combining GLLP \cite{GLLP}, Decoy \cite{Decoy}, and
Recurrence \cite{GVV}, the data post-processing residue rate will be
given by, substituting Eqs.~\eqref{Rec:KVS}, \eqref{Rec:KSV},
\eqref{Rec:KSS}, \eqref{Rec:KSM} and \eqref{Rec:KMS} into
Eq.~\eqref{Rec:Residue1},
\begin{equation} \label{App:Residue2}
\begin{aligned}
r =&  -\frac12f(p_S)H_2(p_S)-\frac12p_Sf(\frac{\delta^2}{p_S})H_2(\frac{\delta^2}{p_S})+K_{VS}+K_{SV}+K_{SS}+K_{SM}+K_{MS} \\
\ge&-\frac12f(p_S)H_2(p_S)-\frac12p_Sf(\frac{\delta^2}{p_S})H_2(\frac{\delta^2}{p_S})\\
& +\Omega_V\Omega\left[\frac14
-\frac14(1-e_1)H_2\left(\frac{e_1-a}{1-e_1}\right)
-\frac14e_1H_2\left(\frac{a}{e_1}\right)\right]
\\
& +
\Omega_V\Omega\left[\frac12-\frac12(1-e_1)H_2\left(\frac{e_1-a}{1-e_1}\right)-\frac12e_1H_2\left(\frac{a}{e_1}\right)\right]
\\
&+ \Omega^2 [1 - e_1(1-e_1) -
\frac12(1-e_1)H_2\left(\frac{e_1-a}{1-e_1}\right)-\frac12e_1H_2\left(\frac{a}{e_1}\right) \\
&-\frac12(1-e_1)^2H_2\left(\frac{e_1-a}{1-e_1}\right)-\frac12e_1^2H_2\left(\frac{a}{e_1}\right)]
\\
& +
\Omega\Omega_M[\frac12-\frac12(1-e_1)H_2\left(\frac{e_1-a}{1-e_1}\right)-\frac12e_1H_2\left(\frac{a}{e_1}\right)]
\\
& + \Omega\Omega_M[\frac12-\frac12e_M(1-e_1)-\frac12e_1(1-e_M)\\
&-\frac12(1-e_1)(1-e_M)H_2\left(\frac{e_1-a}{1-e_1}\right)-\frac12e_1e_MH_2\left(\frac{a}{e_1}\right)]\\
\end{aligned}
\end{equation}
with equality when $q_{11}^V=1/4$ and $q_{11}^M=e_M/2$. In order to
simplify this formula, we define some variables,
\begin{equation} \label{Rec:Const}
\begin{aligned}
B &=
\frac12f(p_S)H_2(p_S)+\frac12p_Sf(\frac{\delta^2}{p_S})H_2(\frac{\delta^2}{p_S})
\\
C &= \frac34\Omega_V\Omega + \Omega^2 (1-e_1+e_1^2) +
\frac12\Omega\Omega_M(2-e_1-e_M+2e_1e_M)
\\
D_1 &=
\frac34\Omega_V\Omega+\frac12\Omega^2(2-e_1)+\frac12\Omega\Omega_M(2-e_M)
\\
D_2 &=
\frac34\Omega_V\Omega+\frac12\Omega^2(1+e_1)+\frac12\Omega\Omega_M(e_M+1)
\end{aligned}
\end{equation}
Thus Eq.~\eqref{Rec:Residue2} can be expressed as
\begin{equation} \label{App:Residue3}
\begin{aligned}
r =& -B+K_{VS}+K_{SV}+K_{SS}+K_{SM}+K_{MS} \\
\ge& -B+C-F_a
\\
\end{aligned}
\end{equation}
where
\begin{equation} \label{Rec:Fab}
\begin{aligned}
F_a&=D_1(1-e_1)H_2(\frac{e_1-a}{1-e_1})+D_2e_1H_2(\frac{a}{e_1}) \\
\end{aligned}
\end{equation}
with equality when $q_{11}^V=1/4$ and $q_{11}^M=e_M/2$.

To lower bound $r$ in Eq.~\eqref{App:Residue3}, we need to find the
maximum value of $F_a$ over the free variable $a$.
We are interested in the range of $a\in[0,e_1]$ with $e_1\le 1/2$.
Note that $F_a$ is a concave function of $a$ in the valid range,
since a sum of two concave functions is also a concave function, and
reflecting and shifting a concave function is also a concave
function. Thus, we can take the derivative of $F_a$ with respect to
$a$ and set it to zero to find the maximum of $F_a$. Differentiating
$F_a$ with respect to $a$ gives
\begin{eqnarray*}
\frac{d F_a}{d a} &=& D_1 \left[ \log_2 \left( \frac{e_1-a}{1-e_1}
\right) - \log_2 \left( 1-\frac{e_1-a}{1-e_1} \right) \right] + D_2
\left[ \log_2 \left( 1-\frac{a}{e_1} \right) - \log_2 \left(
\frac{a}{e_1} \right) \right]
\end{eqnarray*}
Setting $2^{\frac{d F_a}{d a}} = 1$ gives
\begin{eqnarray*}
\left(\frac{1-e_1}{e_1-a}-1\right)^{-D_1}
\left(\frac{e_1}{a}-1\right)^{D_2} &=& 1.
\end{eqnarray*}
Denoting the left-hand side to be $f(a)$, $f(a)$ is a decreasing
function of $a$ since $\frac{d F_a}{d a}$ is a decreasing function
of $a$. Therefore, we can use the bisection method to find $a$ such
that $f(a)=1$. The initial range for the bisection method is
$[0,e_1]$.


\end{appendix}



\begin{thebibliography}{99}
\bibitem{PNS}
Eve can selectively surpass all the single photon signals from
Alice, and split all the multi photon signals, keeping one copy
herself and send the other copy to Bob. In this way, Eve could have
an identical copy of what Bob processes, thus breaking the security
of BB84 protocol.
\end{thebibliography}
\end{document}